\documentclass[aps,prb,twocolumn,superscriptaddress,showpacs]{revtex4}

\usepackage{graphicx, bm, amsmath}

\newcommand{\s}{\hspace{1ex}}
\newcommand{\rb}[1]{\raisebox{2ex}[0pt]{#1}}

\begin{document}

%\preprint{}

\title{Influence of static Jahn-Teller distortion on the magnetic excitation spectrum of PrO$_2$: A synchrotron x-ray and neutron inelastic scattering study}



\author{C.\,H.\ Webster}
\email{carol.webster@npl.co.uk} \affiliation{National Physical
Laboratory, Queens Road, Teddington, Middlesex, TW11 0LW, United
Kingdom}

\author{L.\,M.\ Helme} \author{A.\,T.\ Boothroyd} \affiliation{Clarendon Laboratory,
University of Oxford, Parks Road, Oxford, OX1 3PU, United Kingdom}

\author{D.\,F.\ McMorrow} \affiliation{London Centre for Nanotechnology and Department of
Physics and Astronomy, University College London, London, WC1E
6BT, United Kingdom}

\author{S.\,B.\ Wilkins} \author{C.\ Detlefs}
\affiliation{European Synchrotron Radiation Facility, Bo{\^i}te
Postale 220, F-38043, Grenoble C{\'e}dex, France}
\altaffiliation[Present address: ]{CMPMSD, Brookhaven National
Laboratory, Upton, New York 11973, USA}

\author{B.\ Detlefs}
\affiliation{European Commission, JRC, Institute for Transuranium
Elements, Postfach 2340, Karlsruhe, D-76125, Germany}
\affiliation{European Synchrotron Radiation Facility, Bo{\^i}te
Postale 220, F-38043, Grenoble C{\'e}dex, France}

\author{R.\,I.\ Bewley}
\affiliation{ISIS Facility, Rutherford Appleton Laboratory, Chilton,
Didcot, OX11 0QX, United Kingdom}

\author{M.\,J.\ McKelvy} \affiliation{Center for Solid State
Science, Arizona State University, Tempe, Arizona 85287-1704, USA}

%\thanks{}


\date{\today}

\begin{abstract}
A synchrotron x-ray diffraction study of the crystallographic
structure of PrO$_2$ in the Jahn-Teller distorted phase is reported.
The distortion of the oxygen sublattice, which was previously
ambiguous, is shown to be a chiral structure in which neighbouring
oxygen chains have opposite chiralities.  A temperature dependent
study of the magnetic excitation spectrum, probed by neutron
inelastic scattering, is also reported. Changes in the energies and
relative intensities of the crystal field transitions provide an
insight into the interplay between the static and dynamic
Jahn-Teller effects.
\end{abstract}

\pacs{61.10.Nz, 71.70.Ch, 71.70.Ej, 78.70.Nx}
%61.10.Nz = X-ray diffraction
%71.70.Ch Crystal and ligand fields
%71.70.Ej Spin-orbit coupling, Zeeman and Stark splitting, Jahn-Teller effect
%78.70.Nx Neutron inelastic scattering

%\keywords{}

\maketitle


\section{Introduction}

In recent years there has been strong interest in the interplay
between electric and magnetic degrees of freedom in lanthanide and
actinide dioxides, e.g. UO$_2$, NpO$_2$ and PrO$_2$.\cite{Santini}
These compounds share many similar structural and magnetic
characteristics, but also present a wealth of unusual individual
behaviour.  For example, the longstanding mystery over the nature of
the single phase transition at 25.5 K in NpO$_2$ brought about
suggestions first of octupolar ordering,\cite{Santini:2000, Paixao}
then of triakontadipolar ordering.\cite{Santini:2006} Also, a recent
study of UO$_2$ by resonant x-ray magnetic scattering has revealed
electric quadrupolar ordering that coincides with a Jahn-Teller
distortion and the onset of antiferromagnetic
ordering.\cite{Wilkins:2006} PrO$_2$ exhibits an unusually large
static Jahn-Teller distortion at 120 K,
\cite{Gardiner:PrO2Distortion} as well as a strong dynamic
Jahn-Teller effect due to the high degree of orbital degeneracy in
the ground state.\cite{Boothroyd:2001}  In all three compounds
magnetoelastic interactions play a role in determining the low
temperature properties.  However, due to the complexity of such
interactions, the extent of their contribution is not yet fully
understood.\cite{Bevilacqua, Fournier, Cowley:1968, Sasaki,
Caciuffo:1999}

In this paper we focus on PrO$_2$, which exhibits strong
correlations between its structure and magnetism.  At room
temperature it is paramagnetic and possesses the fluorite crystal
structure. Below $T_{\rm D}$ = 120 K it undergoes a cooperative
Jahn-Teller distortion which doubles the unit cell along one
axis.\cite{Gardiner:PrO2Distortion} Magnetic ordering does not
accompany this transition, but below $T_{\rm D}$ it is possible to
influence the population of structural domains through the
application of a magnetic field.\cite{Gardiner:PrO2MagneticField}
Below $T_{\rm N}$ = 13.5 K the Pr spins order antiferromagnetically,
adopting a magnetic structure with two components: one with the same
unit cell as the fluorite crystallographic structure, and the other
with the same unit cell as the distorted structure below $T_{\rm
D}$.\cite{Gardiner:PrO2Distortion} Below $T_{\rm N}$ the populations
of both structural and magnetic domains can be influenced by the
application of a magnetic field.\cite{Gardiner:PrO2MagneticField}

We present a synchrotron x-ray diffraction study of the
crystallographic structure, which distinguishes between two
previously ambiguous possibilities for the distortion below $T_{\rm
D}$.  Both possibilities consist of an internal distortion of the
oxygen sublattice, but one involves a shear and the other a chiral
deformation.  We also present temperature dependent measurements of
the magnetic excitation spectrum, probed by neutron inelastic
scattering, which show that the dynamic Jahn-Teller effect persists
over a wide range of temperature from 1.8 K to room temperature.
Changes to the crystal field excitations observed on cooling through
the Jahn-Teller transition are interpreted using a point-charge
model of the crystalline electric field at the Pr site, drawing on
the results of the synchrotron x-ray diffraction study. The
implications of our results for existing models of the dynamic
Jahn-Teller effect \cite{Boothroyd:2001, Bevilacqua} are discussed.

\section{X-ray diffraction experiment}

\begin{figure}[!ht]
\begin{center}
\includegraphics{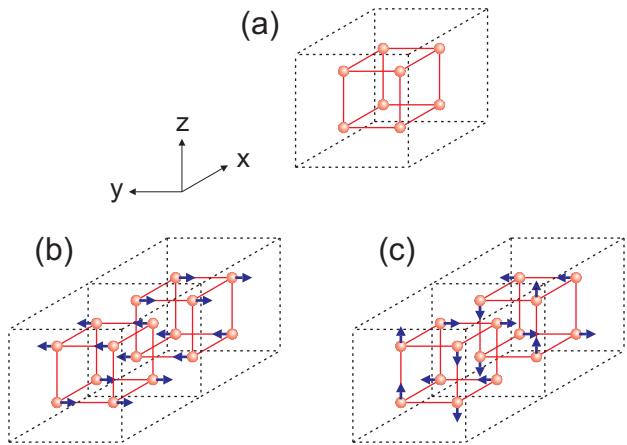}
\caption{(Color online).  (a) PrO$_2$ possesses the fluorite crystal
structure above $T_{\rm D}$.  The spheres are the oxygen ions. The
Pr ions are not shown here, but they occupy positions at the corners
and face centres of the dashed cube. (b) Sheared structure. (c)
Chiral structure. (b) and (c) are alternative models for the
structural distortion that occurs below $T_{\rm D}$. The arrows show
the directions in which the oxygen ions are displaced.}
\label{fig:structures}
\end{center}
\end{figure}

The fluorite structure (space group $Fm\bar{3}m$) exhibited by
PrO$_2$ at room temperature is illustrated in Fig.\
\ref{fig:structures}(a). In a recent neutron diffraction study we
found that a cooperative Jahn-Teller transition at $T_{\rm D}$ = 120
K caused an internal distortion of the oxygen sublattice which
doubled the unit cell along the [100]
axis.\cite{Gardiner:PrO2Distortion} However, in that study the exact
configuration of the oxygen ions remained ambiguous. Two possible
structures were suggested: an orthorhombic structure (space group
$Imcb$), shown in Fig.\ \ref{fig:structures}(b), in which the cubic
oxygen sublattice was sheared along a single crystallographic
direction, and a tetragonal structure [space group $I4(1)/acd$],
shown in Fig.\ \ref{fig:structures}(c), in which the oxygen ions
were displaced chirally, with neighbouring chains possessing
opposite chiralities.

In the paper published simultaneously with this one, Jensen
describes a theoretical model which shows that the chiral structure
is the more stable of the two.\cite{Jensen} Although both structures
exhibit identical structure factors for the Bragg reflections
measured previously by neutron diffraction, a separate set of
reflections exists, which distinguishes the two.  This set is
present for the chiral structure and absent for the sheared
structure. However, the structure factors are extremely weak
compared to those of the parent fluorite structure.  As large single
crystals are not available, we decided to exploit the high dynamic
range achievable with synchrotron x-ray diffraction to study these
weak reflections.

We present the results of an experiment performed at the European
Synchrotron Radiation Facility, Grenoble, France on the instrument
ID20.  The sample used for the experiment was a single crystal of
mass $\sim 1$ mg (the same sample as used for previous neutron
diffraction studies), possessing a single flat face. \footnote{The
flat face of the crystal was close to the (111) plane, but did not
coincide directly with a plane of high symmetry.  Therefore, a Laue
image was used to align the crystal with respect to the instrument.}
The sample was enclosed in a closed-cycle refrigerator, with a base
temperature of 12 K, which was mounted within a five-circle
diffractometer with a vertical scattering plane. An incident energy
of 10 keV was used, corresponding to a wavelength of 1.24 \AA.

We studied three sets of Bragg reflections, two of which were
previously studied by neutron diffraction. The first set,
corresponding to the fluorite structure, obeys the selection rule
$h$, $k$, $l$ all even or all odd.  We label this set $F$. These
reflections contain scattering intensity from both the Pr and O
atoms, but the cross section for x-rays is dominated by the
contribution from Pr due to its much larger atomic number. Since the
Jahn-Teller (JT) distortion is predominantly an internal distortion
of the oxygen atoms, the fluorite reflections remain strong below
$T_{\rm D}$. The second set, corresponding to the distorted
structure below $T_{\rm D}$, and arising purely from oxygen
scattering, obeys the selection rule $h = (2n + 1)/2$, $k$ = odd,
$l$ = even, $l \ne 0$ (or permutations thereof). We label this set
$JT(I)$. This set is common to both the sheared and chiral
structures, so it cannot help us distinguish between the two. The
third set, which also arises purely from oxygen scattering, was not
previously studied by neutron diffraction.  This set obeys the
selection rule $h$ = odd, $k = 2n$, $l = 4n$ (or permutations
thereof), and is present for the chiral structure only.  We label it
$JT(II)$.

\begin{figure}[!ht]
\begin{center}
\includegraphics{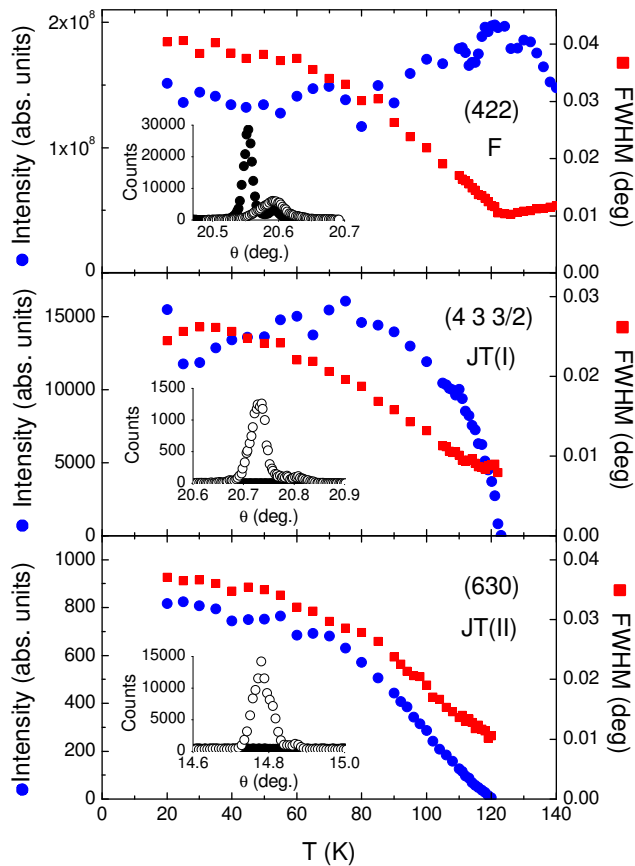}
\caption{(Color online).  Temperature dependence of the integrated
intensity (blue circles) and FWHM (red squares) of typical peaks
from sets $F$, $JT(I)$ and $JT(II)$. The insets show $\theta$-scans
below and above $T_{\rm D}$ (open circles: $T = 20$ K, closed
circles $T \ge 120$ K).  Integrated intensities are normalized to
the incident beam intensity.  The insets show detector counts, which
are not normalized.} \label{fig:theta-scans}
\end{center}
\end{figure}

First, we selected three Bragg reflections typical of each of the
sets described above, and measured the temperature dependence of
these using $\theta$-scans \footnote{A $\theta$-scan in a
five-circle vertical diffractometer is the equivalent of an
$\omega$-scan in a four-circle horizontal diffractometer.  This type
of scan involves the rotation of the crystal about an axis normal to
the scattering plane, while keeping the scattering angle $2\theta$
fixed.} over a range of temperatures between 20 K and 150 K.  A Ge
(111) analyzer crystal was used for these scans in order to optimize
the resolution and dynamic range. The integrated intensity and the
full width at half maximum (FWHM), obtained by fitting a Lorentzian
profile to each $\theta$-scan, are plotted as a function of
temperature for the three peaks in Fig.\ \ref{fig:theta-scans}.
Above $T_{\rm D}$ only the fluorite peak (422) is present.  Below
$T_{\rm D}$, the oxygen peaks $\left(43\frac{3}{2}\right)$ and (630)
appear.  The presence of the $JT(II)$ peak (630) immediately rules
out the sheared structure and provides strong evidence in favour of
the chiral structure.

Next, we measured all the accessible peaks in each set using
$\theta$-scans at 20 K to determine the crystallographic structure
factors.  These scans were done without the Ge (111) analyzer
crystal in order to improve the counting statistics. The integrated
intensities were obtained by fitting Lorentzian profiles and were
subsequently corrected for the geometric Lorentz factor and the
absorption factor.  The peaks in set $F$ were also corrected for
attenuation occurring from an aluminium plate of 1 mm thickness that
was placed in the beam to prevent detector saturation. Finally, the
intensities were multiplied by a universal scaling factor to allow
direct a comparison with the squared structure factors $\vert
F(Q)\vert^2$ calculated for the chiral structure.  The scaling
factor was chosen such that the intensity of the (630) reflection
matched its squared structure factor exactly.

Table \ref{tab:StrucFactors} compares the corrected scaled
intensities of the observed peaks with the calculated squared
structure factors $\vert F(Q)\vert^2$ for the chiral structure,
assuming each oxygen ion is displaced by $0.0726$ \AA\ in the
direction shown in Fig.\
\ref{fig:structures}(c).\cite{Gardiner:PrO2Distortion} The observed
intensities agree with the calculated $\vert F(Q)\vert^2$ to within
a factor of 4 for sets $JT(I)$ and $JT(II)$, and to within an order
of magnitude for set $F$.  Better agreement cannot be expected due
to the difficulty in estimating the absorption factor away from the
specular condition.  There is also an additional uncertainty on the
measured intensity of the peaks in set F due to the possible
nonuniformity of the thickness of the aluminium attenuator.  In
addition to the measurements shown in Table \ref{tab:StrucFactors}
we checked several reflections whose indices did not satisfy the
selection rules for sets $F$ and $JT(II)$.  All of these were
absent, as expected.

\begin{table}[!ht]
\renewcommand{\arraystretch}{1.5}
\begin{center}
\begin{tabular}{|c|c||c|c|}
\hline
Set & Reflection & $\left|F({\bf Q})\right|^2$ (fm$^2$) & Intensity (fm$^2$)\\
\hline \hline
              & $\left(311\right)$ & $2.6 \times 10^4$ & $2.0 \times 10^4$\\
              & $\left(222\right)$ & $1.7 \times 10^4$ & $1.5 \times 10^5$\\
              & $\left(333\right)$ & $1.7 \times 10^4$ & $1.6 \times 10^5$\\
\rb{$F$}      & $\left(422\right)$ & $2.4 \times 10^4$ & $1.3 \times 10^5$\\
              & $\left(402\right)$ & $1.4 \times 10^4$ & $5.3 \times 10^4$\\
              & $\left(531\right)$ & $1.4 \times 10^4$ & $4.1 \times 10^4$\\
              \hline \hline
              & $\left(2\frac{1}{2}1\right)$ & 4.9 & \s 2.5\\
              & $\left(4\frac{1}{2}1\right)$ & 5.5 & \s 2.8\\
$JT(I)$       & $\left(63\frac{3}{2}\right)$ & 4.3 & 13.6\\
              & $\left(43\frac{3}{2}\right)$ & 3.3 & \s 5.9\\
              & $\left(6\frac{1}{2}1\right)$ & 5.3 & 15.0\\
              \hline \hline
              & $\left(630\right)$ & 0.3 \s & 0.3 \s\\
              & $\left(401\right)$ & 0.2 \s & 0.2 \s\\
\rb{$JT(II)$} & $\left(201\right)$ & 0.04   & 0.03\\
              & $\left(432\right)$ & 0.05   & 0.19\\
              \hline
\end{tabular}
\caption{Comparison between the squared structure factors
$\left|F({\bf Q})\right|^2$ predicted for the chiral structure
[Fig.\ \ref{fig:structures}(c)] and the integrated intensities of
the observed x-ray reflections at $T = 20$ K.  $F$ denotes fluorite
reflections, $JT(I)$ denotes reflections from the oxygen
displacements that are common to both the sheared and chiral
structures, and $JT(II)$ denotes reflections arising solely from the
chiral structure.  Uncertainties are not given for the integrated
intensities, as these are dominated by the uncertainty in the
absorption factor, which is difficult to quantify.}
\label{tab:StrucFactors}
\end{center}
\end{table}

\section{Neutron scattering experiment}

In a previous experiment we probed the low temperature excitation
spectrum of PrO$_2$ using neutron inelastic scattering on a powder
sample.\cite{Boothroyd:2001}  The spectrum showed two main features:
sharp peaks characteristic of crystal field transitions at energies
above 100 meV, and a broad band of scattering in the range 10--100
meV which we interpreted as magnetic scattering from vibronic
excitations associated with the dynamic Jahn-Teller effect.  We
developed a simple model of the magnetoelastic coupling between a
single phonon mode and a crystal field excitation, which yielded a
qualitative agreement with our results.  However, at the time we
were unaware of the existence of a static Jahn-Teller distortion,
and therefore chose a phonon mode consistent with cubic symmetry for
our model. Other researchers have also used cubic symmetry as the
basis for more complex multiphonon models.\cite{Bevilacqua}  In the
light of our present knowledge, that the tetragonal distortion
involves only an internal displacement of the oxygen atoms, leaving
the cubic Pr lattice undisturbed, the choice of cubic symmetry is
not unreasonable.  However, the distortion lowers the crystal field
symmetry at the Pr site and the large magnitude of the oxygen
displacements is expected to cause a significant splitting of the
cubic $\Gamma_8$ ground state, as well as an altering of the
positions and intensities of the higher energy crystal field
transitions.  An examination of the temperature dependence of the
excitation spectrum is therefore in order.

We report the results of a neutron inelastic scattering experiment
performed at the ISIS spallation source, Didcot, UK on the high
energy transfer (HET) time-of-flight chopper spectrometer.  A powder
sample of PrO$_2$ was prepared from a starting material of
Pr$_6$O$_{11}$, which was baked at 1000 $^{\circ}$C for several
hours, then annealed in flowing oxygen at 280 $^{\circ}$C for $\sim$
5 days. The products were checked by x-ray diffraction and no trace
of residual Pr$_6$O$_{11}$ could be found.  A sample of mass 25.83 g
was sealed in an aluminium foil packet and mounted inside a
temperature-controlled top-loading closed-cycle refrigerator.  The
spectrum was measured at 7 K, 50 K, 80 K, 100 K, 110 K, 115 K, 120
K, 130 K, 165 K and 200 K, using two different incident energies at
each temperature: $E_{\rm i}$ = 80 meV (chopper frequency 350 Hz)
and $E_{\rm i}$ = 250 meV (chopper frequency 500 Hz). An identical
set of measurements (although at a reduced set of temperatures: 7 K,
80 K, 120 K and 165 K) was made on a sample of CeO$_2$ of mass 25.98
g, sealed in a similar foil packet, to estimate the non-magnetic
background due to phonons and multiple scattering in
PrO$_2$.\footnote{CeO$_2$ possesses the fluorite structure and has a
similar lattice parameter to PrO$_2$. The neutron scattering lengths
of Ce and Pr differ by no more than 10\%, but the Ce$^{4+}$ ion has
no $4f$-electrons, making it non-magnetic. Hence, CeO$_2$ provides a
good estimate of the non-magnetic scattering from PrO$_2$.} Vanadium
spectra were used to calibrate the detectors and convert the
scattering intensity into absolute units of cross-section, i.e.\
mb\,sr$^{-1}$\,meV$^{-1}$\,(Pr ion)$^{-1}$.

\begin{figure}[!ht]
\begin{center}
\includegraphics{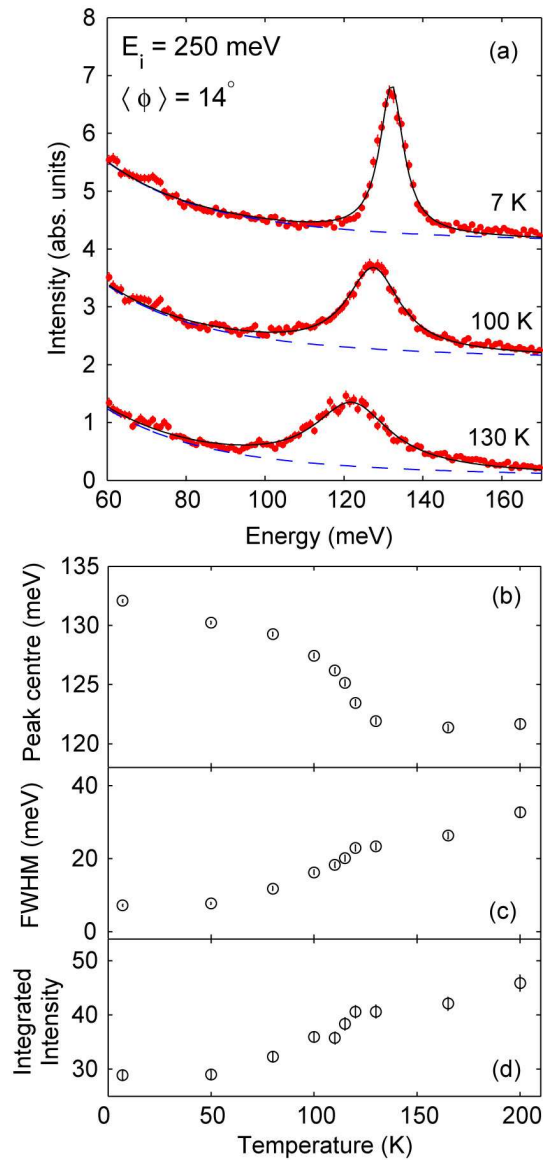}
\caption{(Color online).  (a) Crystal field transition from the
ground state to the $\Gamma_7$ excited state in PrO$_2$, measured by
neutron spectroscopy.  The red circles are data from the low angle
detector banks of the HET spectrometer. Data taken at different
temperatures have been offset vertically for clarity. The solid
black line is a fit to the data using a Lorentzian profile on a
background estimated from the CeO$_2$ data (dashed blue line). (b),
(c) and (d) show the temperature dependence of the peak centre, the
FWHM and the integrated intensity, respectively.} \label{fig:Gamma7}
\end{center}
\end{figure}

Figure \ref{fig:Gamma7}(a) shows PrO$_2$ spectra measured in the low
angle detector banks with an incident energy of $E_{\rm i}$ = 250
meV, corrected for absorption and self-shielding.  The single peak
is due to a crystal field transition from the ground state to the
$\Gamma_7$ excited state. Above $T_{\rm D}$ the peak is centred at
121 meV, but as the sample is cooled through the Jahn-Teller
transition the width of the peak decreases sharply and its centre
moves rapidly to higher energies, reaching 132 meV at 7 K.  The data
are fitted using a Lorentzian profile on a background estimated from
the CeO$_2$ data. The fit parameters are used to plot the
temperature dependence of the peak centre, FWHM and integrated
intensity in Fig.\ \ref{fig:Gamma7}(b).

The change in energy of the peak below $T_{\rm D}$ can be understood
qualitatively by constructing a simple model of the crystalline
electric field, ignoring the spatial extent of the electron
wavefunctions and treating the Pr$^{4+}$ and O$^{2-}$ ions as point
charges.  This type of model underestimates the separation of the
crystal field levels because it disregards the overlap between the
electron orbitals.  However, it gives a reasonable idea of the
relative energy separation of the levels and also their relative
spectral weights.  If the separation of a pair of levels is known
experimentally, the point-charge model can be scaled to match, thus
enabling us to predict the approximate positions of other levels.

\begin{figure}[!ht]
\begin{center}
\includegraphics{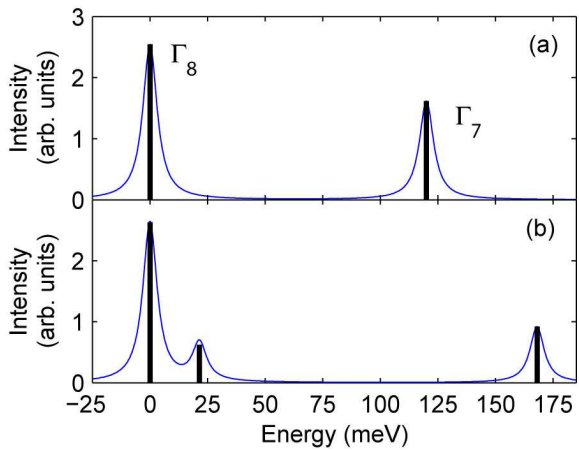}
\caption{(Color online).  (a) Point-charge model for the fluorite
crystal structure above $T_{\rm D}$.  The model has been scaled to
match the observed $\Gamma_8\rightarrow\Gamma_7$ transition energy
of 121 meV. The thick black vertical lines represent the peak
intensities.  The thin blue line is a simulation of the neutron
scattering excitation spectrum, but is not in absolute units of
cross section. (b) Point charge model for the chiral structure below
$T_{\rm D}$, using the same scaling factor as the fluorite model.}
\label{fig:PointCharge}
\end{center}
\end{figure}

Figure \ref{fig:PointCharge}(a) shows a point-charge simulation of
the excitation spectrum for the fluorite phase of PrO$_2$ above
$T_{\rm D}$.  The simulation takes into account the resolution
function of the HET spectrometer at an incident energy of 250 meV
and assumes a sample temperature of $T = 0$ K, but the intensity is
not normalized to absolute units of cross section. In the fluorite
phase, the ground state is a $\Gamma_8$ quartet and the excited
state is a $\Gamma_7$ doublet. The separation of these has been
scaled to match the observed transition energy of 121 meV. Figure
\ref{fig:PointCharge}(b) shows the simulated spectrum from a point
charge model of the chiral phase below $T_{\rm D}$, using the same
scaling factor as for the fluorite phase.  It is immediately obvious
that the $\Gamma_7$ level has moved to a higher energy, in agreement
with our experimental observations (although the magnitude of the
observed shift on cooling through $T_{\rm D}$ is rather less than
the point-charge prediction). The model also shows that the spectral
weight of the $\Gamma_7$ level decreases in the chiral phase, which
accounts for the discrepancy between model and experiment in our
previous study.\cite{Boothroyd:2001} Lastly, the $\Gamma_8$ ground
state is predicted to split into two doublets, separated by $\sim
21$ meV.

\begin{figure}[!ht]
\begin{center}
\includegraphics{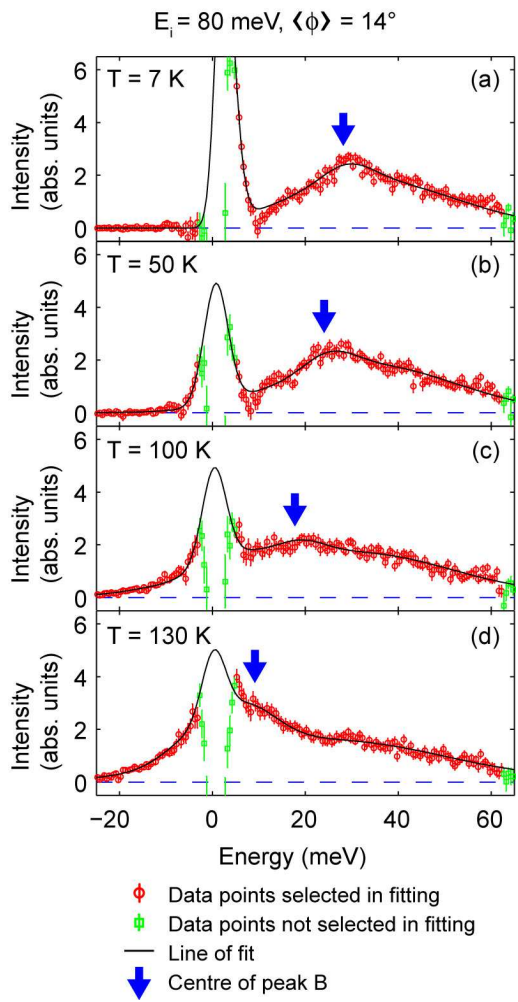}
\caption{(Color online).  (a)--(d) PrO$_2$ spectra with the CeO$_2$
background subtracted (circles), showing the variation of the broad
band of vibronic scattering with temperature.  The black line is a
fit, constructed from four separate peaks (see Fig.\
\ref{fig:Fits}), which is used to determine the temperature
dependence of the crystal field transition between the split
doublets of the $\Gamma_8$ ground state in the chiral phase.  The
centre of the transition, determined from the fit, is indicated in
each plot by a blue arrow.} \label{fig:Vibronic}
\end{center}
\end{figure}

The results of the point-charge model lead us to examine the low
energy spectrum of PrO$_2$, to see if we can detect the splitting of
the $\Gamma_8$ ground state below $T_{\rm D}$. Figure
\ref{fig:Vibronic} shows the difference between the PrO$_2$ and
CeO$_2$ spectra measured in the low angle detector banks with an
incident energy of $E_{\rm i}$ = 80 meV. CeO$_2$ data for
temperatures intermediate between actual measurements were estimated
by interpolation.  The broad band of scattering above 10 meV is
clearly present both below and above $T_{\rm D}$. We observe that
the scattering above 35 meV is relatively independent of
temperature.  However, the region below 35 meV is strongly
temperature dependent with a maximum that shifts from $\sim$ 28 meV
at 7 K to $\sim$ 20 meV at 100 K, before becoming quasielastic close
to $T_{\rm D}$.  The energy of the maximum at 7 K is in good
agreement with the prediction of the point-charge model for the
chiral phase and also with other predictions\cite{Kern:1964, Jensen}
for the splitting of the $\Gamma_8$ ground state in the distorted
phase below $T_{\rm D}$.  This provides strong evidence that the
broad band of scattering contains a component arising from a crystal
field transition between the split doublets of the $\Gamma_8$ ground
state.

\begin{figure}[!ht]
\begin{center}
\includegraphics{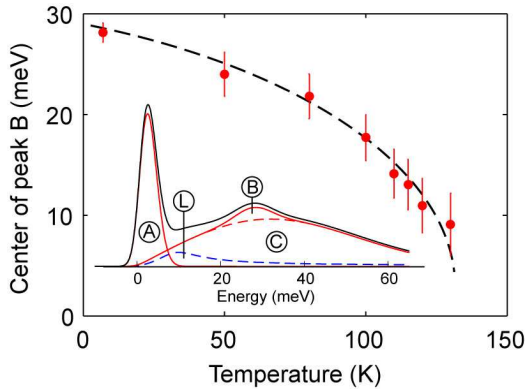}
\caption{(Color online).  Temperature dependence of the ground state
splitting in the chiral phase. The circles are the results of the
fit to all spectra between 7 K and 130 K, where the centre of peak B
was allowed to vary with temperature. The dashed line is a guide to
the eye. The inset shows the functions used to construct the
lineshape which was fitted to the data. A, B and C are symmetrized
Gaussians, whereas L is a symmetrized Lorentzian. All are weighted
by a detailed balance factor.} \label{fig:Fits}
\end{center}
\end{figure}

In order to extract the temperature dependence of the splitting in
an unbiased way, we constructed a lineshape to fit the broad band of
scattering from three symmetrized Gaussians and a symmetrized
Lorentzian, as shown in Fig.\ \ref{fig:Fits}. Peak A represents
quasielastic scattering from the ground state, peak B represents the
transition between the two doublets of the split $\Gamma_8$ ground
state, and peak C represents a broad continuum of vibronic
scattering.  Peak L (the symmetrized Lorentzian) was added to
improve the quality of the fit at low energies.  Each peak was
weighted by a detailed balance factor to take into account the
thermal population of the levels.\cite{Helme}  In order to determine
the width and amplitude of peak B, the lineshape was initially
fitted to the spectrum at 7 K.  We then fitted the lineshape
simultaneously to all the spectra for temperatures between 7 K and
130 K, keeping the width and amplitude of peak B fixed. The other
peak widths, centres and amplitudes were free parameters, but were
not allowed to vary with temperature.  The only parameters allowed
to vary with temperature were the centre of peak B and the amplitude
of peak L.

The solid black lines in Fig.\ \ref{fig:Vibronic} show the fitted
lineshape, which agrees well with the data.  Figure \ref{fig:Fits}
shows the temperature dependence of peak B.  The peak clearly moves
to lower energies as the temperature rises, becoming quasielastic
near $T_{\rm D}$.

\section{Summary and Conclusions}

We have presented the results of two experiments on PrO$_2$: a
synchrotron x-ray diffraction experiment which confirms that the
displacement of the oxygen atoms in the Jahn-Teller distorted phase
is chiral and a neutron inelastic scattering experiment which probes
the crystal field transitions above and below the Jahn-Teller
transition temperature $T_{\rm D}$. The cubic $\Gamma_8$ ground
state is found to split into two doublets in the distorted phase,
and the splitting is found to be $\sim$ 20--30 meV, in good
agreement with a simple point charge model of the crystal field.

Our results indicate that existing models of the dynamic Jahn-Teller
effect \cite{Boothroyd:2001, Bevilacqua} should be revised to
include the effect of the static distortion.  The accompanying paper
by Jensen \cite{Jensen} presents such a model, which exhibits an
improved agreement between the calculated and measured intensities
of the crystal field transitions.  Our neutron scattering data show
that vibronic excitations due to the dynamic Jahn-Teller effect are
largely independent of temperature, despite the presence of a static
Jahn-Teller transition.  It is likely that this is because the
magnetoelastic energy\cite{Boothroyd:2001, Jensen} is comparable to
the crystal field splitting of the ground state in the distorted
phase. Hence, the static Jahn-Teller effect does not quench the
dynamic Jahn-Teller fluctuations.

\vfill

\begin{acknowledgments}
We would like to thank J. Jensen for many useful discussions and D.
Prabhakaran for help with the preparation of the powder sample of
PrO$_2$. Financial support for L.M.H. by the EPSRC is also
acknowledged.
\end{acknowledgments}

\bibliography{PrO2PrBCO}

\end{document}